# Empirical Study of Gaze Behavior in Children and Young Adults Using Deep Neural Networks and Robot Implementation: A Comparative Analysis of Social Situations

**Ramtin Tabatabaei[1], Milad Hosseini[1], Ali Mohajerzarrinkelk[1], Ali F. Meghdari[1,2], Alireza Taheri[1,*]**

[1] Social and Cognitive Robotics Laboratory, Center of Excellence in Design, Robotics, and Automation (CEDRA), Sharif University of Technology, Tehran, Iran

[2] Chancellor, Fereshtegaan International Branch, Islamic Azad University, Tehran, Iran

[*]**Corresponding Author: artaheri@sharif.edu , Tel: +982166165531**

**Abstract**

People's eye direction indicates their interests and intentions in social interactions. Hence, to be appropriately engaged in social situations, social robots should display proper behavior by directing their gaze to the right person and predicting their actions. In a preliminary exploratory study, our goal was to train deep neural network models to (try to logically) mimic children's and/or adults' gaze behavior in certain social situations to reach this objective. Additionally, we aim to identify potential differences in gaze behavior between these two age groups based on our participants' gaze data. Furthermore, we aimed to assess the practical effectiveness of our adult and children models by deploying them on a Nao robot in real-life settings. To achieve this, we first created two video clips, one animation and one live-action, to depict some social situations. Using an eye-tracking device, we collected eye-tracking data from 24 participants, including 12 children and 12 adults. Then, we utilized deep neural networks, specifically Long Short-Term Memory (LSTM) and Transformer Networks, to analyze and model the gaze patterns of each group of participants. Our results indicate that when the models attempted to predict people's locations (in the next frame), they had an accuracy in the range of 62%-70% with one attempt, which increased by ~20% when attempted twice (i.e. the two highest-ranked predicted labels as outputs). As expected, the result underscores that gaze behavior is not a wholly unique phenomenon. We obtained feedback from 57 new participants to evaluate the robot's functionality. These participants were asked to watch two videos of the robot's performance in each mode and then complete a comprehensive questionnaire. The questionnaire results indicate that the participants expressed satisfaction with the robot's interaction, including its attention, intelligence, and responsiveness to human actions. However, they did not perceive the robot as a social companion comparable to a human. This exploratory study tries to address/show potentials of the social acceptance of robots based on human nonverbal behavioral cues for future research.

**Keywords:**

Robot eye gaze; Multiparty interaction; Non-verbal communication; Long Short-Term Memory; Transformers

# 1. INTRODUCTION

Psychology research has demonstrated the criticality of nonverbal behaviors, such as gaze direction, for facilitating turn-taking, social interactions, and revealing hidden mental states, including cognitive effort, making it a crucial nonverbal cue for social communication [1-3]. Nonverbal social cues, including changes in gaze direction, are relied upon by humans to predict intentions [4] and are essential for conveying meaning during social interactions [3]. As machines, including humanoid robots, become increasingly prevalent as assistants, educators, and companions in social and workplace settings like homes, workplaces, hospitals, and schools [5], understanding human communication and nonverbal behaviors, particularly those related to eye gaze, is critical [6, 7]. In response to this trend, research on human-robot interaction (HRI) is increasingly focused on how robots can be perceived as socially intelligent, considering factors such as their actions, speech, and gestures [8]. The involvement of social robots in society is on the rise [9], and fortunately, they are becoming more capable of comprehending both verbal and nonverbal communication, which is essential for them to interact effectively with humans [5]. These nonverbal cues play a crucial role in collaborative tasks requiring close physical proximity, ensuring high efficiency and preventing misunderstandings [7]. Realistic gaze behaviors are critical for creating believable real/artificial agents that can indicate attention and engagement with users by looking in a particular direction, which plays an essential role in human-robot communication [2, 10]. Gaze models can help agents to interact more effectively with humans by carrying the agent's internal state, which can be inferred by either humans or other agents [10].

Research in gaze behavior can be classified into three distinct categories: human perception and interaction with robots, human-robot communication, and social interaction with ethical considerations.

In the field of human perception and interaction with robots, Perugia et al. [11] measured participants' gaze patterns and perceptions of robots and found participants demonstrated an aversion to gazing at the robot in social chats, indicating an uncanny valley effect, and the more the participants gazed at the robot during a joint task, the worse their performance. Koller et al. [12] conducted a pilot study to determine the influence of gaze aversion ratio (GAR) on human interaction experience (IE) with a social robot. They found that a longer GAR leads to a positive IE. Kshirsagar et al. [13] studied robot gaze behavior during human-robot handovers. They found that a transition gaze from the giver's face to their hand is perceived as more likable, anthropomorphic, and communicative of timing. Mutlu et al. [14] explored how a humanoid robot's gaze can be modeled for storytelling. Their results showed that participants performed better in recalling the story when the robot looked at them more and that there were gender differences in how the robot ASIMO was evaluated based on gaze frequency. Kamelabad et al. [15] analyzed how a social robot's appearance and interaction impact language learning. Their results showed that children learned better and viewed the robot as more human-like when playing alone compared to in pairs. The robot's appearance as an adult or child did not affect learning outcomes. Xu et al. [16] examined how people reacted to quadruped robots in different scenarios and showed that the robot's orientation and gaze affect the distance people prefer to keep from the robot, indicating their attitudes towards it. Tatarian et al. [17] examined how different behaviors of social robots, such as body language and speech, affected their perceived social intelligence. They found that using proxemics, gaze mechanisms, kinesics, and social dialogue all had varying impacts on the robots' perceived social intelligence.

In the field of Human-robot communication, Cumbal et al. [18] created an adaptive robot to assist adults in learning a second language. The robot modified its dialogue and feedback based on the learner's proficiency and motivation, resulting in increased speaking time, fluency, and improved proficiency. Aliasghari et al. [19] analyzed how a trainee robot's nonverbal cues impact human teachers' perception of its learning behavior. They found that effective nonverbal communication design, including the robot's gaze behavior and arm movement trajectory, is essential to improve learning outcomes and social interactions with human teachers. Lathuilière et al. [20] proposed a neural network-based reinforcement learning approach for autonomous robot gaze control in human-robot interaction. The experimental evaluation showed that their method is effective and robust, with the best results obtained using both audio and visual information. D. Domingo et al. [21] explored two models for improving robot gaze control in human-robot interaction: a competitive neural network and a recurrent neural network with LSTM layers. Duque-Domingo et al. [22] improved robots' focus on specific interlocutors using a neural network method during face-to-face communication, tested on a robotic head, the method achieved human-like behavior and handled attention transitions smoothly. Ise et al. [23] examined if tactile perception could replace visual perception for recognizing a robot while walking. Walking hand in hand with the robot caused participants to look around longer than walking alongside it.

In the field of Social interaction and ethical considerations, Pasquali et al. [24] developed a method for a robot to detect lies during human-robot interactions based on pupil dilation. They conducted an experiment utilizing the iCub humanoid robot during a card game and obtained promising results as the robot successfully detected deception. Lyu et al. [25] presented a method for mobile robots to announce their upcoming operation to pedestrians using a rotating 3D face interface. The experiments showed that the method effectively reduces pedestrian discomfort during robot encounters. Lehmann et al. [7] examined how people interact with a Nao robot positioned at eye level in personal and intimate spaces. They found that participants did not fully scale down their personal space for the robot's size, and the leaning-back behavior of the robot was sometimes perceived as unfriendly. Schellen et al. [26] found that establishing eye contact during human-robot interaction in response to deceptive behavior increased participants' honesty. Kompatsiari et al. [27] investigated how users evaluated a humanoid robot based on whether it established eye contact or not. Their results showed that eye contact led to higher engagement and perceived human likeness. Babel et al. [28] examined how a social robot's communication affects trust and found that different strategies work better for different dialogue content. A directed gaze was better for small talk, and robot initiative increased trust during the first service task interaction.

To the best of our knowledge, there is a limited number of studies utilizing deep neural networks to model human gaze behavior in social situations, particularly in the context of multi-party conversations. Furthermore, there is a lack of research on the gaze behaviors of children and young adults in the existing literature (, separately). This paper aims to address this gap and presents our primary contribution. This preliminary exploratory study presents an empirical motion-time pattern for human gaze behaviors (including children and young adults) in diverse social situations. The objectives of the paper are as follows: 1) Train models to logically mimic human gaze behavior in various social situations with reasonable performance, 2) Compare the effectiveness of models derived from both the children and adults while also examining the impact of animated versus live-action videos in data collection, 3) Compare different features of adults and children while they watch these video clips to understand potential variations in gaze behavior, and 4) Investigate the performance of the trained models when implemented on a social robot during multi-person social scenarios.

## 2. METHODOLOGY

This section is organized into three main parts. Firstly, it provides a detailed exploration of the experiment setup and participant information. It covers the different designed video clips, their corresponding features, and the social situations depicted in the clips. Furthermore, it discusses the selection process for participants and the gaze capture setup. Secondly, the section focuses on the architectural design of the neural networks, including specifications for the input and output, as well as the various layers involved. Lastly, it details the implementation of these models onto a robot and outlines the necessary setup requirements. Additionally, it explains the evaluation process used to assess the robot's performance and provides a comprehensive overview of the assessment criteria.

### 2.1. Experimental Setup and the Participants

Our goal is to model and compare the gaze direction of children and adults in some common social situations. We also aim to investigate whether gaze behavior differs between participants watching live-action videos versus animation videos. In this regard, two video clips were prepared to collect eye-tracking data from human participants. Both videos have a resolution of 1920 × 1080 pixels and a duration of 2 minutes, for a total of 3,600 frames, recorded at a speed of 30 frames per second. The videos feature different scenarios unfolding with two to four individuals and a fixed box. In the animated video, the individuals are designed as animated men with identical faces and clothing. In contrast, the individuals in the live-action video include two girls and two boys; although their presence or absence may vary in different scenarios, the individuals are the same throughout the video. In both videos, the box remains fixed in place and does not move. The described scenario involves four people arranged in different orientations (-60, -30, 30, and 60 degrees, respectively) with respect to the user's viewpoint. These individuals are standing and engaged in various activities, and their angles remain fixed unless they are walking (entering or leaving the scene), in which case their angles can change. Additionally, there is a box located directly in front of the user at an angle of 0 degrees. Each person may exhibit different characteristics such as presence/absence, proximity to the user, right-hand pointing, left-hand waving, talking, and entering/exiting.

A total of 24 different scenarios are conducted, with each scenario lasting approximately 5 seconds before smoothly transitioning to the next mode. It is important to note that these transitions occur continuously rather than in a discrete manner, ensuring a seamless flow between the different scenarios. This controlled environment enables researchers to systematically manipulate different parameters and observe people's responses to various stimuli, thus mitigating the confounding effects of uncontrolled factors. **Figure 1-a** displays a segment of a live-action video clip at the 98[th] second. In the footage, person number 1 is situated in the distance near the observer and is not engaging in any of the mentioned activities. Person number 2 is located far from the observer and is talking, pointing, and waving. Person number 3 has the same characteristics as person number 1. Person number 4 is far from the observer and is pointing at a box. **Figure 1-b** shows a segment of an animated video clip at the 38th second. Person number 1 is situated in the distance near the observer and is not engaging in any of the mentioned activities. Person number 2 is not present, while person number 3 is located nearby and is waving and pointing. Person number 4, similar to person number 2, is not present.

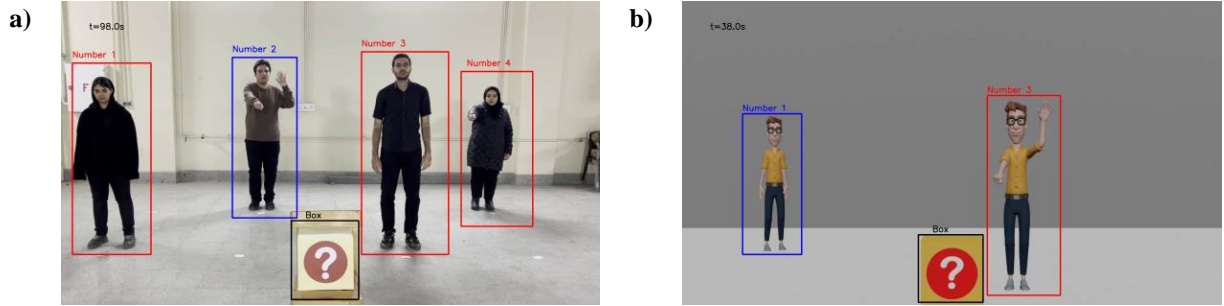

**Figure 1.** Positions of the people and the box in **a)** the live-action video and **b)** the animated video.

The SR Research EyeLink 1000 Plus device was utilized to conduct the eye-tracking test at the Mowafaghian Comprehensive Rehabilitation Center in Tehran, Iran. This advanced eye-tracking device has a data capture rate of up to 2000 Hz and a measurement error of only approximately 0.15 degrees, which ensures that the data obtained from the test is both precise and dependable. Additionally, the device is equipped with a head holder, which ensures that the participants' head is securely positioned during the examination, minimizing the possibility of unwanted movements and ensuring that the captured data is uniform and consistent. The experimental setup can be seen in **Figure 2**. Utilizing this cutting-edge technology guarantees that the results obtained from the test are of high quality and accurately reflect the visual behavior of the participants.

Two groups of participants were involved in the experiment: children (12 boys) with an average age of 7.13 years old and a standard deviation of 0.14 years and adults (including six men and six women) with an average age of 21.66 years old and a standard deviation of 0.94 years. All participants were without any apparent physical abnormalities to ensure the representativeness of the results.

During the experiment, participants viewed an animation and a live-action video, each lasting for 120 seconds, while the SR Research EyeLink 1000 Plus device captured data at a frequency of 1000 Hz. This resulted in a substantial amount of data, with approximately 120,000 x and y values for each participant. A sliding window approach was utilized to assign x and y values to each frame, averaging the data points over three consecutive frames: the first and second frames were generated by averaging the first 33 data points, while the third frame was obtained by averaging the last 34 data points within each window. Due to factors such as blinking, head movement, and device calibration issues, it is possible that certain data may be missing or fall outside the expected range. As a result, we purposely excluded such data and their corresponding frames from our neural network design to ensure the accuracy of the experiment's results.

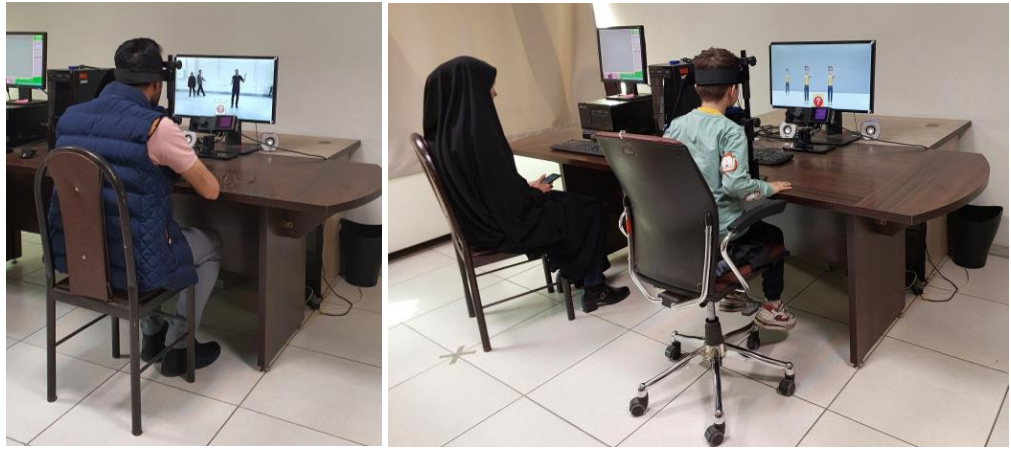

**Figure 2.** The eye-tracking experimental setup.

**2.2. Extracting Empirical Motion-Time Gaze Pattern using Deep Neural Networks**

The aim of this section is to design models to predict where the robot should look based on past social situations. Two deep neural network models were employed for this task: LSTM and Transformers. Both neural networks were given input in the format of (n, 30, 28), where n represents the total number of the data points, 30 denotes the frame sequence size, and 28 signifies the total number of features in each frame. The total number of features can be visualized as 4×7 matrices. The dimensionality of 4 indicates the overall count of individuals in the video, whereas the dimensionality of 7 signifies the number of significant attributes each person can potentially display. The matrix contains data for each individual in the video clip, with each row representing a different person. The data includes several characteristics, including whether the person is present in the scene (indicated by 0 or 1), the distance between the person and the observer (measured in meters), whether the person is waving (0 or 1), whether the person is pointing (0 or 1), whether the person is speaking (0 or 1), the angle at which the person is facing the observer (measured in degrees), and the person's movement status (standing, entering, or exiting the scene, where 0 indicates standing, 1 indicates entering, and 2 indicates leaving). Sensors, such as the Kinect in this study, can easily provide the necessary inputs for social robots. These sensors detect the positions of landmarks on each person's body and hands, and the kinematic energy is computed by analyzing these positions to determine various characteristics.

To prepare the input for both networks, we start by identifying the characteristics of each individual in every frame. These characteristics are initially discrete within each frame, but when formed into sequences of characteristics, the social situations transition smoothly from one social situation to another for the input of the neural networks. Furthermore, we tracked the position of their right hand when pointing and their left hand when waving, as depicted in **Figure 3**. Our initial experiment involves five classification labels: "Person 1," "Person 2," "Person 3," "Person 4," and the "Box." However, in our second experiment, we distinguished between the body and hands, resulting in 13 classification labels, with an additional eight labels specifically for hand gestures. We labeled each data frame with the person ID/box the participant was looking at, which served as the basis for our gaze control system's predictions and detections. Specifically, we aimed for our system to accurately predict the label and choose its gaze direction based on the previous frames of the video.

In order to ensure the validity of our data, we excluded instances where the participants did not look at any of the people or the box after the events sequence had occurred, as they (non-human-based situations) were considered/assumed to be noise data. Using this methodology, we labeled the remaining frame sequences for both video clips across all 24 participants. By carefully filtering out irrelevant data, we obtained a more accurate and reliable dataset for our analysis.

Our study aimed to investigate differences in the gaze behavior of children and adults while watching animated and live-action videos. To accomplish this, we extracted eight different models from our dataset. The first four models were generated by analyzing the data of children and adults watching each type of video separately. The fifth and sixth models were created using merged data from both age groups to determine if video type influenced the accuracy of the models. Finally, the seventh and eighth models were based on merged data from animated and live-action videos to compare the accuracy of the models trained based on the children's and adults' gaze directions. The accuracy of the models measures how well the trained models mimic human behavior in the same situation, taking into account a few preceding frames. It also reflects the similarity of label detection among individuals in social situations. To train the models, we used a frame sequence size of 30 (1 second) with a step size of 1 (for the moving window on the time-series data). We labeled each sequence frame with the next frame output label indicating the location where the model should detect to look at. The reason for choosing a frame sequence size of 30 with a step size of 1 refers to another of our current works [29] in which we concluded that the frame sequence size of 1s (30 frames in this study) is the optimal choice as a larger frame sequence size can lead to delayed predictions.

We used the Keras library in Python to implement our two deep models, i.e., LSTM and transformers, using a categorical cross-entropy loss function, an Adam optimizer, and early stopping at 10 based on the testing data's accuracy, with a maximum of 100 epochs, and a batch size of 20. Our dataset consisted of the gaze data from the 24 participants. To ensure the robustness of our models, we evaluated their accuracy using K-Fold cross-validation, with K set to 6 for all models. In other words, we trained each model six times, using five-sixths of the participants' data for training and one-sixth for testing. The average accuracy of the six training runs was then reported for each model. To facilitate our training process, we utilized Google Colab. The following subsections provide a detailed account of the specific architecture used in each approach.

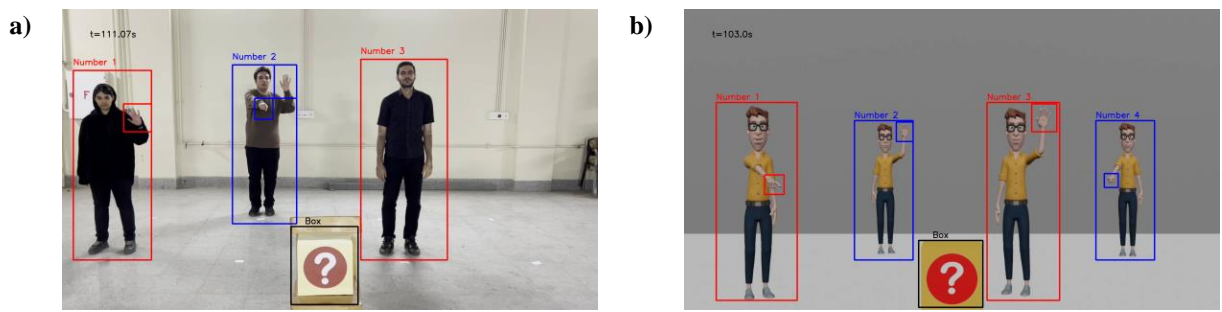

**Figure 3.** Labels to be detected by the model in **a)** the live-action video and **b)** the animated video.

### 2.2.1. Modelling with the LSTM neural network design

The LSTM architecture is a powerful tool for capturing temporal dependencies in sequential data. LSTM is specifically designed to handle sequential data where the input has a temporal or time-based dimension. LSTM is well-suited for capturing long-term dependencies and patterns in sequential data by selectively retaining or forgetting information over a longer period of time. LSTM units consist of several components, including a memory cell, input gate, forget gate, and output gate, controlled by sigmoid and tanh activation functions. In our proposed model architecture, we utilized two layers of LSTM, with each layer consisting of 64 units and employing a hyperbolic tangent activation function (tanh). The output of the second LSTM layer was directed towards a softmax layer, which contained either 5 or 13 neurons for classification, depending on the specific experiment. The input data was passed through the two LSTM layers before being processed by the output layer, as illustrated in **Figure 4**.

The dimensions of the training data varied slightly depending on the age group of the participants or the type of video they were watching (live-action or animation). This is because missing data is removed from the training data for each experiment. For example, in one experiment using live-action movie data from adults, the input array has dimensions of (31772, 30, 28). In this case, the first dimension (31772) represents the number of data instances used for training after removing missing data at certain time intervals. The second dimension (30) represents the sequence size, which refers to the number of frames (or snapshots) included in each sequence. The third dimension (28) denotes the concatenation of all rows (the participants' characteristics in the mentioned frame). When using 13 labels, the number of training parameters for this particular model is 57,677. When using 5 labels, it is 51,157. The number of training parameters refers to the number of adjustable parameters in the model that are used to minimize the error between the predicted and actual labels.

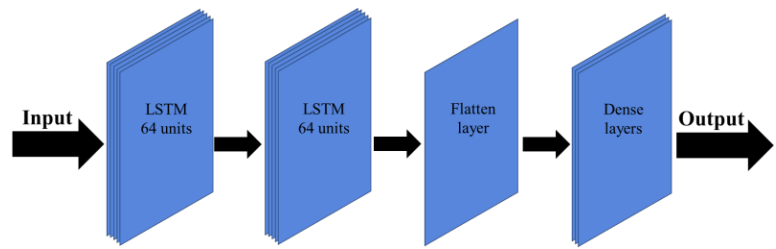

**Figure 4.** LSTM network architecture.

## 2.2.2. Modelling with the Transformers neural network design

The transformer architecture has been a groundbreaking innovation in natural language processing, employing self-attention mechanisms to process input sequences. It can capture long-range dependencies more effectively than traditional recurrent neural networks by simultaneously attending to all input tokens. Our system takes the input as normalized numbers in vectors, eliminating the need for Input Embedding. It employs positional encoding with an input dimension of 30 (representing the size of the frame sequence) and an output dimension of 28 (representing the total number of features). Our system utilizes a single Encoder block consisting of a MultiHeadAttention component with two attention heads. The size of each attention head was set to 28 (matching the total number of features). Additionally, the Encoder block includes a position-wise feed-forward network with two dense layers using the swish activation function. The dimensions of these layers are 1024 and 28, respectively. An element-wise addition operation was performed to combine the outputs of the Encoder block with the original inputs. Subsequently, the resulting vector underwent GlobalMaxPooling1D, which extracted the maximum value from each feature dimension, reducing the sequence length. Finally, the processed vector was passed through a dense layer with the dimension equal to the number of labels (e.g., 5 or 13), generating the final predictions. The model architecture is depicted in **Figure 5**. The attention mechanism highlights the importance of input tokens, which are then processed through multiple dense layers to generate the final output predictions.

The transformer network's input data (and its dimension/shape) is exactly the same as our LSTM network. For a model using 13 labels, the number of training parameters is 66,193, and for 5 labels, it is 65,961.

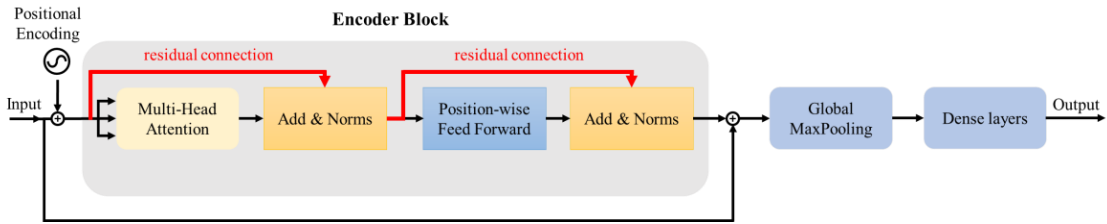

**Figure 5.** The Transformers network architecture.

### 2.2.3. Implementation and Evaluation of gaze models on the social robot

In this section, we discuss how we implemented our machine-learning models onto a Nao robot. To evaluate and compare the real-world performance of our models (the adult and children models), we asked participants to complete a questionnaire and rate their attitudes and feelings towards the robot using the Glasgow Coma Scale. In addition, we investigated the participants' attitudes regarding their satisfaction with the robot's interaction, human-like qualities of the robot, intelligence of the robot, and understanding of human actions.

In order to accurately determine the position, sound direction, and action of individuals in front of the robot, we employed the use of the advanced Kinect 2 technology. With its cutting-edge capabilities, Kinect 2 is able to detect the total number of individuals present and provide the precise position of each person's landmark. By establishing specific thresholds and utilizing kinematics energy, we are able to detect the actions and movements of each individual. Additionally, Kinect 2's high-quality microphone arrays allow for the detection of the direction of incoming sound. Overall, the combination of Kinect 2's advanced features and the sophisticated neural network designs enables our robot to better understand and respond to the people in its surroundings.

The implemented network operated on the Nao robot. The Nao is widely regarded as one of the most aesthetically pleasing humanoid robots, with its sleek white design and rounded features giving it a friendly and approachable appearance. Developed by Softbank Robotics in France, the Nao is a result of years of research and development in the field of (social) robotics, and it has become a popular choice for use in various robotic groups worldwide. One of the Nao's notable abilities is its ability to rotate its head at different speeds, which allows for greater flexibility in its interactions with humans, enabling it to track movements more accurately and respond more quickly to changes in its environment. This feature, combined with its other impressive capabilities, makes the Nao robot a highly versatile and effective tool for a wide range of applications, from education and healthcare to entertainment and beyond.

To evaluate the real-world performance of our model, we provided two 2-minute videos of the robot performances based on the children/adults' gaze patterns in a set of social situation scenarios. We conducted a study involving 57 new participants who were asked to view the videos (via social media) and complete a questionnaire assessing their emotions and attitudes toward the robot and its performances (some other details are also presented in the results section, subsection 5). The questionnaire utilized the Glasgow Coma Scale to gauge participants' responses in two distinct scenarios: one involving the best model trained with the children and the other with the young adults. An anonymous questionnaire was developed, encompassing variables such as age, gender, and a set of questions inspired by the UTAUT [30] and the questionnaire used in reference [31] to assess the robot's performance. The participants were asked to indicate their level of agreement with each of the ten statements presented in **Table 1**. A five-point Likert scale was utilized to rate the robot's social presence. The scale included verbal anchors ranging from 'totally disagree' to 'totally agree', allowing participants to express their opinions on the statements. To ensure impartiality and minimize potential bias, the order of the videos and statements was randomized before conducting the assessment (counterbalance condition). This shuffling of the videos/statements aimed to create an unbiased evaluation environment.

**Table 1.** The evaluation questions to assess the robot's performance.

| *Statements* | |
|---|---|
| Statement 1: | I feel satisfied with the interaction of the robot in the video. |
| Statement 2: | I'm satisfied with the robot's coordination with people's movements in the video. |
| Statement 3: | In my opinion, the robot demonstrates a good understanding of the people in the video. |
| Statement 4: | I believe the robot in the video is a great social companion. |
| Statement 5: | The robot's interaction with the people in the video felt remarkably human-like to me. |
| Statement 6: | The robot in the video appears so lifelike that I can easily imagine it as a living being. |
| Statement 7: | The robot in the video pays good attention to the people around it. |
| Statement 8: | The robot behaved intelligently in the video. |
| Statement 9: | The robot in the video responded adeptly to the actions of the people. |
| Statement 10: | The robot in the video exhibited a solid understanding of the people's actions. |

## 3. RESULTS

In this section, we present and analyze the results of the two neural network designs (LSTM and transformers) used in the study. The accuracy of each model, which was trained using datasets from both children and adults, is compared in the first three subsections. The accuracy of the models shows how closely the trained models imitate human behavior in similar situations. Two cases are considered: one where the entire body of each person is utilized, excluding their hands and heads, and another where the body and hands are treated as separate labels. In subsection four, we discuss the p-value obtained from a two-sample t-test conducted on the dataset to determine whether there were any significant differences in the gaze behaviors between children and adults. Finally, in the fifth subsection, the best model obtained from the children's and adults' datasets was implemented on the Nao robot, and some new participants were asked to rate the robot's performance in real-world scenarios.

### 3.1. Comparative Analysis of the LSTM and Transformer Models for Gaze Prediction in Children and Adults Watching Animated and Live-Action Movies

**Figure 6** illustrates the accuracy of the two neural network architectures used for labeling individuals in two ways: one by labeling individuals alone and the other by labeling the individual and the location of their hands while performing specific activities. The graphs display the accuracy of the models based on the number of detection attempts, which provides the models with an increased opportunity to identify the labels correctly.

**Figure 6-a** and **Figure 6-b** depict the accuracy of the models and the likeness of the models' emulation of human gaze behavior in social situations. In particular, the first figure focuses on the accuracy of the models when labeling each person individually. In contrast, the second figure highlights the accuracy when labeling both the person and the specific box they occasionally pointed to. The results demonstrate that the models achieved the highest scores when analyzing gaze behavior during the children's viewing of animated movies. On the other hand, the lowest accuracy occurred when adults watched real-life movies, with LSTM-based models achieving accuracies of 69.6% (±2.7%) and 61.6% (±9.4%), respectively. The two graphs illustrate that the model's accuracy improves as the number of detection attempts increases. In the case of children watching the animated film, the model achieves 88.4% (±2.0%) accuracy after two attempts, which further increases to 96.5% (±1.3%) after three attempts. Interestingly, the models achieved higher accuracy when trained with children's datasets. Notably, both the LSTM and Transformers neural network architectures demonstrated similar results in terms of accuracy, with an accuracy difference of less than 0.82%.

**Figure 6-c** and **Figure 6-d** present the outcomes of the models, depicting the labeling of each person's body, including their hand locations and the corresponding pointed boxes in certain instances. Remarkably, the models exhibit only marginal variations in accuracy. The highest accuracy was achieved when the models were tested on children and adults watching the animated film. In contrast, the lowest accuracy was observed when adults watched the live-action film. The LSTM and Transformers neural network architectures exhibited remarkable similarity in their results, with an accuracy difference of less than 0.58%. The accuracy of the LSTM model differed for different viewer groups. When children watched the animated movie, the model achieved an accuracy of 64.9% (±2.8%), whereas, for adults watching the live-action movie, the accuracy was 59.1% (±8.5%). Both graphs provide evidence that the similarity in correctly identifying the same label increases as the number of recognition attempts increases. Specifically, after two attempts, the accuracy for the children watching the animated film reached 81.0% (±2.8%), which further improved to 89.4% (±2.0%) after three attempts.

The LSTM models trained with both children and adults datasets were evaluated to assess their similarity scores in predicting labels for social situations in live-action and animated movies. In the live-action domain, the models with 13 labels achieved an average similarity of 77.94% (± 3.1%), while the 5-label models attained 72.97% (± 3.3%). For animated movies, the models with 13 labels achieved an average similarity of 74.87% (± 3.0%), compared to 75.42% (± 2.5%) for the 5-label models. The Transformers models show results almost similar to those of the LSTM. Each condition had 6 different models due to the K-fold setting, resulting in a total of 36 model comparisons. The average and standard deviation of these 36 comparisons were reported by comparing every set of six models with their corresponding counterparts. These findings indicate that, despite similar accuracy percentages in the models trained with adult and children datasets, there can be differences in the labels predicted by the models in the same social situations. This discrepancy may suggest different gaze behavior between children and adults.

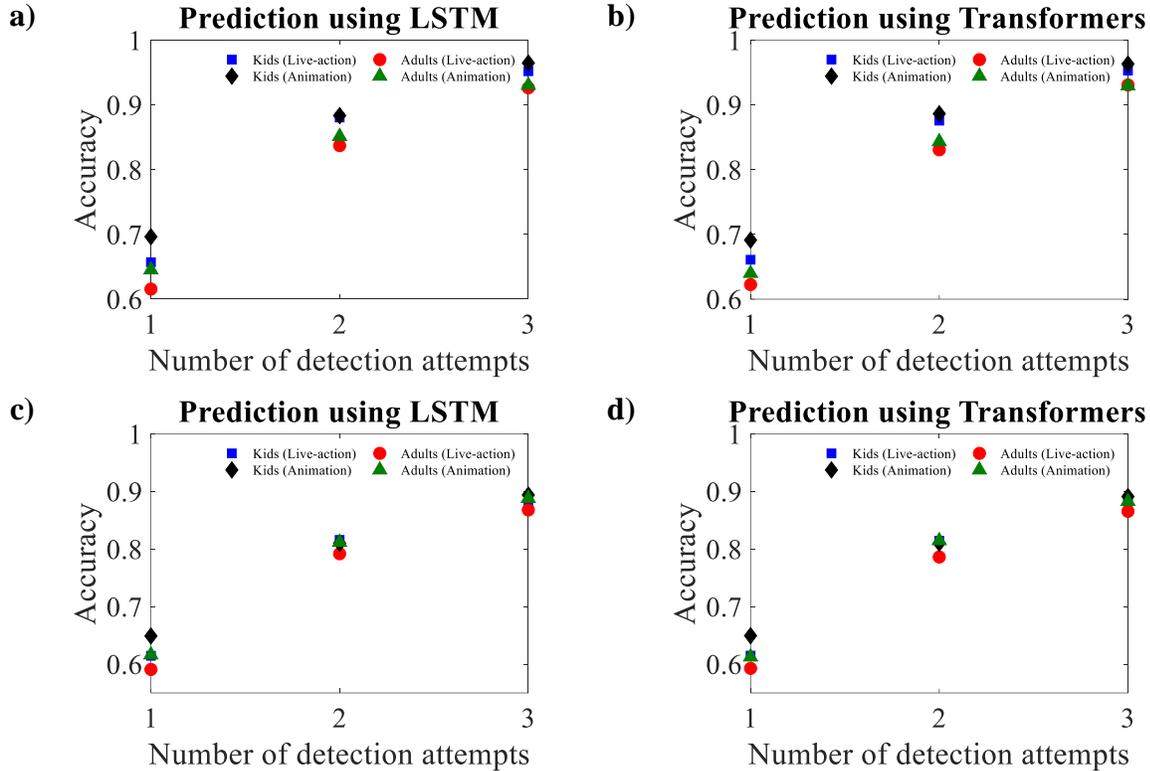

**Figure 6.** Comparison of the accuracy of four models using LSTM and Transformers neural network designs. Parts **(a)** and **(b)** show the accuracy of LSTM and Transformers neural network designs with five labels (four persons and the box). Similarly, parts **(c)** and **(d)** display the accuracy of the LSTM and Transformers neural network designs with 13 labels (four persons while considering the location of their hands, heads, and the box).

## 3.2. Comparative Analysis of the LSTM and Transformer Models for Gaze Prediction in Merged Children and Adults Datasets Watching Animated and Live-Action Movies

**Figure 7** displays the accuracy of the two neural network architectures trained with the data from the children and adults, measuring the proximity of the models' imitation to human gaze behavior in similar situations. This section aims to investigate whether the type of film (animation or live-action) impacts the accuracy of the models.

**Figure 7-a** presents the results of the models that label each person and the corresponding box, while **Figure 7-b** showcases the outcomes of models that represent each person's body and hand location with the box. Notably, there was only a slight variation in accuracy between the two graphs, and no significant differences were observed in the performance of the models in each detection attempt. Both LSTM and Transformers neural network architectures yielded similar results, with an accuracy difference of less than 0.51%. Both graphs demonstrate that the accuracy of the models improves with an increase in the number of attempts. In Figure 5, the accuracy reaches 65.6% (±5.5%) with one detection attempt in the animation, which then improves to 87.1% (±4.8%) after two detection attempts. Similarly, in Figure 6, the accuracy starts at 61.7% (±6.6%) with one detection attempt in the animation, and it improves to 81.4% (±6.3%) after two detection attempts.

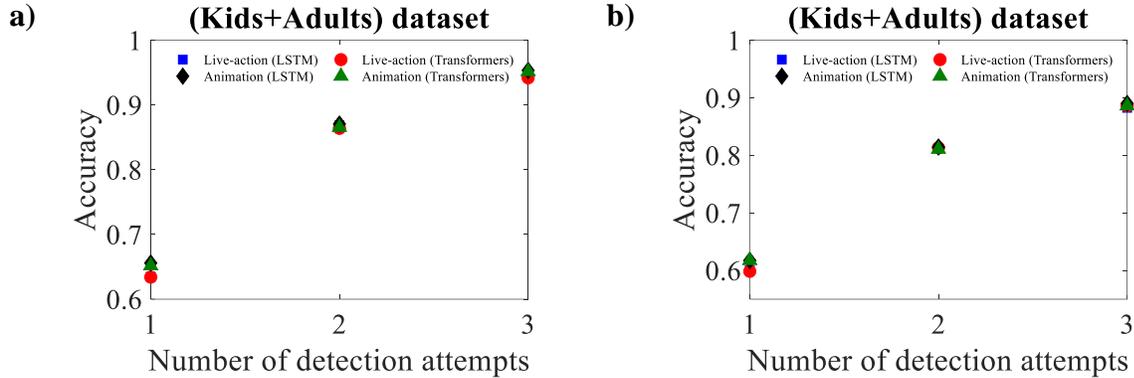

**Figure 7.** Comparison of the accuracy of two models trained with the merged datasets of children and adults using LSTM and Transformers neural network designs. Part **(a)** shows the accuracy of LSTM and Transformers neural network designs with five labels (four persons and the box). Similarly, part **(b)** displays the accuracy of the LSTM and Transformers neural network designs with 13 labels (four persons while considering the location of their hands, heads, and the box).

### 3.3. Comparative Analysis of the LSTM and Transformer Models for Gaze Prediction on Merged Animated and Live-Action Datasets of Children and Adults

**Figure 8** depicts the accuracy of two neural network architectures trained with a merged dataset of animation and live-action, showcasing the degree to which the models replicate human behavior in social situations. The primary objective of this study is to investigate whether the age group (the children or adults) influences the accuracy of these models.

The results are presented in **Figure 8-a** and **Figure 8-b**, highlighting the model outcomes labeled for each individual and box location in **Figure 8-a** and additionally incorporating their hands' location in **Figure 8-b**. Notably, LSTM and Transformers neural network architectures yielded similar results in both scenarios (the accuracy difference is less than 0.36%), with only marginal discrepancies in accuracy between the two datasets. Interestingly, the models trained on the children's dataset exhibited slightly higher accuracy than those trained on the adult dataset. Both graphs illustrate that the model's accuracy improves as the number of detection attempts increases. In **Figure 8-a**, the accuracy for children with one attempt was 65.6% (±5.5%), which increased to 87.1% (±4.8%) after two attempts. Similarly, in **Figure 8-b**, the accuracy for children with one attempt was 61.7% (±6.6%), which improved to 81.5% (±6.3%) after two attempts.

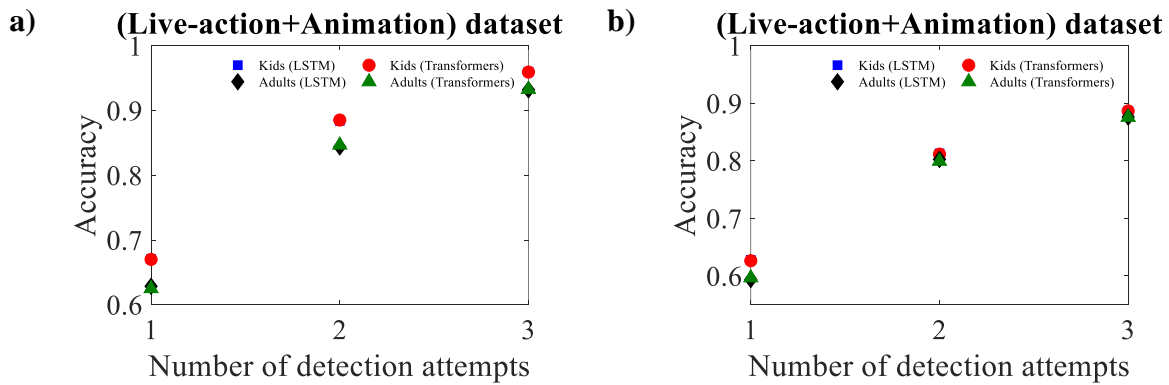

**Figure 8.** Comparison of the accuracy of the two models trained with the merged datasets of the live-action and animation movies using LSTM and Transformers neural network designs. Part **(a)** shows the accuracy of the LSTM and Transformers neural network designs with five labels (four persons and the box). Similarly, part **(b)** displays the accuracy of the LSTM and Transformers neural network designs with 13 labels (four persons while considering the location of their hands, heads, and the box).

### 3.4. A t-test Comparison of the Gaze Behavior between Children and Adults Watching Animated and Live-Action Movies

A two-sample t-test was conducted on three different features, namely 1) the maximum number of frames participants spent looking at a label among 13 labels in a sequence, 2) the total number of times participants shifted their gaze to a different label, and 3) the average number of frames that participants looked at one label. The test was conducted with a confidence level of 95%, hypothesizing a difference of 0 and an alternative hypothesis of Difference $\neq$ Hypothesized difference for the results of the first experiment, which involved 5 labels. Four different comparisons were made; the first two examined whether there is a significant difference between the means of the animation and real-action datasets for both the children and adults. The other two comparisons investigated whether there is a statistically significant difference in means between the children and adults for each of the mentioned three features.

**Table 2** displays the outcomes of a two-sample t-test performed on the animation and real-action datasets. The table exhibits the maximum number of frames during which participants gazed at a label among the 5 labels in a sequence (Max Label Frame), the total number of times participants shifted their gaze to a different label (Total Label Shifts), and the mean number of frames that participants looked at one label (Mean Label Time). The p-values for all features are above 0.05 for adults and children, indicating no significant difference between the two groups at a 95% confidence level. **Table 3** presents the results of a two-sample t-test conducted on the datasets of children and adults. The table shows the Max Label Frame, Total Label Shifts, and Mean Label Time for animation and live-action videos. The p-values for Total Label Shifts and Mean Label Time in both the animation and live-action categories are all below 0.05. The results also indicate that children exhibit a higher number of Total Label Shifts, whereas adults have a higher Mean Label Frame. These findings indicate that there is a significant statistical difference between the two groups in terms of the examined features, which agrees with the findings of [32] and provides empirical evidence that supports the assertion that there are notable disparities in gaze behavior between adults and children. The study revealed a marked difference in the level of inter-subject consistency, with adults demonstrating a higher degree of consistency in their eye movements compared to infants.

**Table 2** Two-sample t-test results for animation and live-action, presented separately for children and adults.

|  | Children | | | Adults | | |
|---|---|---|---|---|---|---|
|  | Animation Mean (SD) | Live-action Mean (SD) | p-Value | Animation Mean (SD) | Live-action Mean (SD) | p-Value |
| Max Label Frame | 203.3 (79.6) | 263.0 (171.0) | 0.291 | 286.0 (119.0) | 248.0 (132.0) | 0.475 |
| Total Label Shifts | 108.2 (22.2) | 103.2 (25.3) | 0.612 | 79.0 (31.6) | 78.5 (33.7) | 0.970 |
| Mean Label Frame | 26.4 (7.1) | 30.0 (5.8) | 0.265 | 44.9 (20.6) | 49.0 (28.9) | 0.691 |

Table 3 Two-sample t-test results for children and adults, presented separately for Animation and live-action (p-values less than 0.05 are shown in bold).

| | Animation | | | Live-action | | |
|---|---|---|---|---|---|---|
| | Children Mean (SD) | Adults Mean (SD) | P-Value | Children Mean (SD) | Adults Mean (SD) | P-Value |
| Max Label Frame | 203.3 (79.6) | 286.0 (119.0) | 0.060 | 263.0 (171.0) | 248.0 (132.0) | 0.819 |
| Total Label Shifts | 108.2 (22.2) | 79.0 (31.6) | **0.017** | 103.2 (25.3) | 78.5 (33.7) | **0.046** |
| Mean Label Frame | 26.4 (7.1) | 44.9 (20.6) | **0.012** | 30.0 (8.0) | 49.0 (28.9) | **0.048** |

**3.5. Implementation and Performance Evaluation of the Best Neural Network Model on a Robotic System**

In the following section, we describe the implementation of our models on the Nao robot using the data from both children and adults datasets. Specifically, we focus on models with the highest accuracy and which take into account the location of each person without considering the location of their hands. For this purpose, we utilized the best LSTM model. We present the participants' evaluations of the robot's performance in real-world settings, discussing their feedback and responses regarding their satisfaction with the robot's interaction, the human-like qualities of the robot, the intelligence of the robot, and the understanding of human actions in further detail in the subsequent sections.

As mentioned earlier, the Nao robot was selected as the ideal choice for implementing the selected models. This decision was primarily based on the robot's ease of use and visually appealing design. However, it is worth noting that the Nao robot lacks a depth sensor. To overcome this limitation, we incorporated the Kinect 2 device, which serves as a reliable depth sensor and also has the capability to detect landmarks of individuals in its field of view. Additionally, the Kinect 2 is equipped with an array of microphones, enabling it to accurately detect the direction of sound. In **Figure 9**, the Nao robot with the integrated Kinect is demonstrated successfully identifying landmarks of the people in front of it.

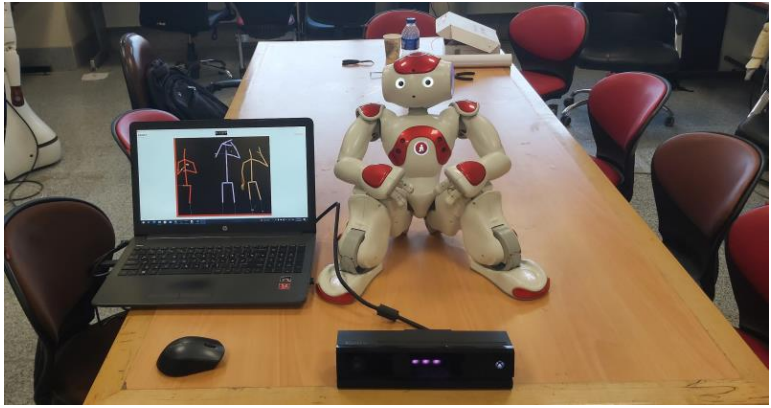

**Figure 9.** Nao Robot Utilizing Kinect for Precise Person Identification.

In order to assess the performance of the models in the Nao robot, we captured two comparable 2-minute videos. Due to the Kinect's limited field of view, the number of individuals standing in front of Nao varied between one and two in our experimental setup. These individuals engaged in activities such as entering, standing, moving, waving, and leaving. To enhance the participants' enjoyment while observing these interactions, the Nao robot greeted the individuals by saying "Hello" when they waved in front of it. Next, these two videos were presented to a group of 57 participants via social media. The participants in this section of the study consisted of 44 males and 13 females, with an average age of 24.98 years old and a standard deviation of 4.20. These participants were distinct from those who underwent the eye-tracking tests. After watching the videos, the participants were then requested to assess the performance of each video by rating their agreement with the statements provided in **Table 1** on a 5-point scale ranging from 'totally agree' to 'totally disagree'. The order of the videos and the questions were shuffled prior to the assessment (the counterbalance condition) to prevent any potential bias. It should be noted that the participants did not have any information about which video referred to the children's/adults' gaze patterns.

A paired t-test with a confidence level of 95% was conducted on the questionnaire results, as shown in **Table 4**. The results indicate that statements 1 and 2, which are related to overall satisfaction and perception of the robot's interaction, received scores higher than 4 in both models. These scores suggest that participants leaned towards agreement, ranging from 'agree' to 'totally agree' with these statements. Statements 4, 5, and 6, which relate to human-like qualities and the robot's realism, received scores around 3. This shows that participants had a neutral stance on these statements, and interestingly, the model trained with adults achieved slightly higher scores. These scores indicate that participants found it challenging to accept the robot as a social companion or view it as human. Statements 7 and 8, concerning the attention and intelligence of the robot, received scores slightly below 4, suggesting that participants were mostly in agreement with these statements. Additionally, participants believed that the model trained with adults demonstrated greater intelligence. Finally, statements 3, 9, and 10, which pertain to the responsiveness and understanding of human actions, received scores similar to those related to attention and intelligence, slightly below 4. These scores indicate that participants were mostly in agreement with these statements and believed that both models exhibited good responsiveness and understanding of human actions. When comparing the scores of children and adults presented in **Table 4** with those obtained in [31], we observed that our models performed significantly better in Attention,

Intelligence, Responsiveness, and Understanding human actions. However, we obtained similar results for human-like qualities and the robot's realism.

**Figure 10-a** displays a frame from the recorded video of the implemented model trained with the adults' dataset on the Nao robot, capturing two individuals in close proximity. Person 2 can be seen waving, and the Nao robot directs its head toward that person. In **Figure 10-b**, a frame from the recorded video showcases the implementation of the model trained with the children's dataset, showing persons 1 and 3 both standing, with person 1 closer to the camera than person 3. Nao's gaze is on person 1 in this frame.

**Table 4** Results obtained from the questionnaire assessing the performance of the children and adult models on the Nao robot. (The p-values that are less than 0.05 are shown in bold)

| Statement number | Children Mean (SD) | Adults Mean (SD) | t-Value | p-Value |
| --- | --- | --- | --- | --- |
| Statement 1 | 4.05 (0.71) | 4.16 (0.56) | -1.03 | 0.31 |
| Statement 2 | 4.21 (0.58) | 4.16 (0.70) | 0.49 | 0.63 |
| Statement 3 | 3.81 (0.93) | 3.93 (0.86) | -1.1 | 0.28 |
| Statement 4 | 3.14 (1.05) | 3.19 (1.05) | -0.5 | 0.62 |
| Statement 5 | 2.86 (1.10) | 3.16 (1.12) | -2.18 | **0.03** |
| Statement 6 | 2.81 (1.02) | 3.04 (1.08) | -2.35 | **0.02** |
| Statement 7 | 3.95 (0.78) | 4.18 (0.70) | -2.43 | **0.02** |
| Statement 8 | 3.53 (0.84) | 3.79 (0.81) | -2.51 | **0.01** |
| Statement 9 | 3.95 (0.66) | 4.12 (0.68) | -1.87 | 0.07 |
| Statement 10 | 3.84 (0.79) | 3.89 (0.77) | -0.48 | 0.64 |

a) 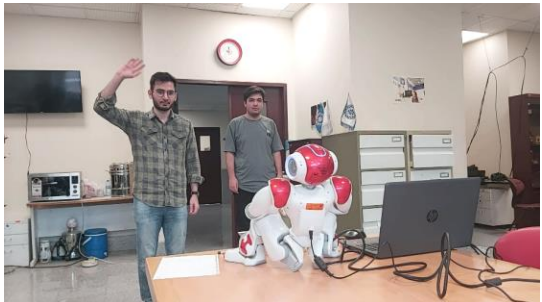 b) 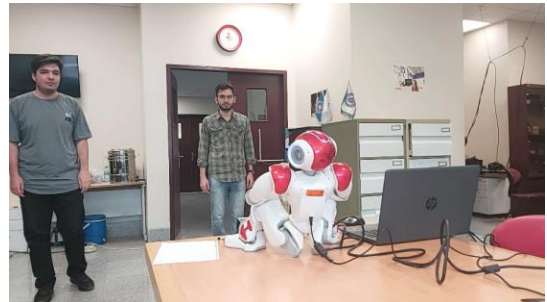

**Figure 10.** A frame of the recorded video of the model's performance on the Nao robot trained with **a)** the adults' dataset and **b)** the children's dataset.

We attempted to find similar studies in this field to compare if our findings are in line with them. However, there was a lack of similar studies proposing gaze models for children/adults implemented on social robots. Hence, we were unable to systematically compare the findings of this research with related studies. More studies need to be conducted in this area across the social robotics community to have more generalizable results.

## 5. LIMITATIONS AND FUTURE WORK

Our study has encountered several limitations that must be considered while interpreting the results. Firstly, the data were collected while the participants' heads were fixed in the eye-tracking device, which caused discomfort and noise in some recordings. Secondly, the use of 2D video clips to collect data may not accurately reflect the complexity of the real-world environment. The utilization of 3D video clips could enhance accuracy. Thirdly, the sample size was limited to only 12 children and 12 adults, and thus, may not be representative of the entire population. Additionally, it's important to acknowledge that our models were trained on a limited set of social situations, which may not fully represent the wide range of diverse social scenarios in the real world. Therefore, we cannot definitively claim that these models are proficient in handling all social situations (however, the outputs are fairly reasonable). Consequently, future studies should include a larger sample size of participants from different age groups and genders to improve the generalizability of the results. Being aware of these limitations is crucial in interpreting the results, and their impact should be considered when applying the findings to real-world situations.

Our future work will focus on enhancing the realism of our data by incorporating additional (both human-based and non-human-based) social situations into our video clips. This approach aims to capture a more extensive range of human behaviors and responses, enabling our model to adapt more effectively to real-world scenarios. Additionally, we plan to explore the utilization of virtual reality technology, along with eye-tracking capabilities, to create a more immersive experience for the participants. These innovations will enable us to capture the subtleties of human gaze behavior in a broader range of social situations, leading to a more reliable and precise model.

## 6. CONCLUSIONS

This preliminary exploratory study involved the use of LSTM and Transformer neural network designs to extract empirical motion-time patterns for a gaze control system of both children and adults. Our objective was to measure how well the trained models mimic human gaze behavior in the same social situations based on the gathered gaze data. Simultaneously, we aimed to identify if there are differences in the gaze behavior of children and adults. Lastly, we aimed to implement our models onto a Nao robot and evaluate their performance in real-life settings. We evaluated the accuracy of the model for both age groups while they were viewing animated and live-action movies. The results indicate that the models achieve an accuracy of approximately 65% when considering the location of each person and the box. However, when incorporating the location of each person's hands, the accuracy slightly decreases to around 62%. This observation holds for both neural network designs after a single detection attempt. Increasing the number of detection attempts to 2 resulted in a ~20% rise in accuracy. This finding underscores that gaze behavior is not a wholly unique phenomenon. However, with giving more chances to the models to detect the correct label, it may be possible to model people's gaze with up to 90% accuracy. Increasing the number of labels from 5 to 13 enhanced the model's capabilities, despite a 6% decrease in accuracy.

In addition, we conducted a comparison of the LSTM models' predictive similarity between the two age groups in terms of assigning labels to social situations. Surprisingly, we discovered 75% similarity between the predictions, despite both groups achieving similar levels of accuracy. After conducting a two-sample t-test on both the animated and live-action datasets, we found no significant difference between the two groups for both age groups at a 95% confidence level. However, when we compared the datasets of children and adults using a two-sample t-test, we observed the children have a higher tendency to shift their gaze from one label to another, with this number being 1.3 times greater than the total number of label shifts observed in adults. On the other hand, adults tend to fixate on a label for a longer duration compared to children, approximately 1.6 times longer. The questionnaire results indicate that the participants expressed satisfaction with the robot's interaction, intelligence, and responsiveness to human actions. However, they held the belief that the robot could not be considered a social companion or be perceived as human-like.

The participants' gaze patterns were influenced by various factors that impacted the locations they tended to look at. These factors encompassed the inherent noise present in human vision, the constraints imposed by the two-dimensional nature of the clips, sensor noise, variations in lighting conditions, as well as individual preferences. Moreover, determining the appropriate gaze direction is not always straightforward, as it can vary due to cultural disparities and individual preferences. Our models have yielded valuable insights into the empirical motion-time pattern of social robot gaze control. Future research should address these challenges by exploring novel solutions to mitigate data noise and enhance the overall accuracy of the gaze control system.

**Acknowledgment**

This study was funded by the "Dr. Ali Akbar Siassi Memorial Research Grant Award" and the Sharif University of Technology (Grant No. G980517).

**Conflict of interest**

Author Alireza Taheri has received a research grant from the Sharif University of Technology (Grant No. G980517). The authors Ramtin Tabatabei, Milad Hosseini, Ali Mohajerzarrinkelk, and Ali F. Meghdari assert that they have no conflict of interest.

**Availability of data and material (data transparency)**

All data from this study are available in the Social & Cognitive Robotics Laboratory archive.

**Code availability:**

All the codes are available in the Social & Cognitive Robotics Laboratory archive. If the readers need the codes, they may contact the corresponding author.

**Authors' contributions:**

All authors contributed to the study's conception and design. Alireza Taheri presented the idea for the study. Ramtin Tabatabaei significantly contributed to the study by designing neural network architectures and implementing them using Python for data analysis and modeling purposes. Additionally, Ramtin Tabatabaei was involved in material preparation and utilized the Nao robot. Milad Hosseini and Ali Mohajerzarrinkelk were responsible for the data collection. Alireza Taheri and Ali F. Meghdari supervised the study. The first draft of the manuscript was written by Ramtin Tabatabaei, and all authors provided comments on previous versions. Finally, all authors read and approved the final manuscript.

**Consent to participate:**

Informed consent was obtained from all individual participants included in the study.

**Consent for publication:**

All participants have consented to submit the results/figures of this study to the journal.